\documentclass[superscriptaddress,pre,twocolumn]{revtex4-2}
\usepackage{amsmath}
\usepackage{amssymb}
\usepackage{amscd}
\usepackage{graphicx}
\usepackage{bbm}
\usepackage{dsfont}
\usepackage{datetime}
\longdate

\begin{document}
\title{Colloids in Two-Dimensional Active Nematics: Conformal Cogs and Controllable Spontaneous Rotation}
\author{Alexander J.H. Houston}
\affiliation{School of Mathematics, University of Leeds, Leeds, LS2 9JT, United Kingdom}
\affiliation{School of Physics, Engineering and Technology, University of York, Heslington, York, YO10 5DD, United Kingdom}
\author{Gareth P. Alexander}
\email{G.P.Alexander@warwick.ac.uk}
\affiliation{Department of Physics, Gibbet Hill Road, University of Warwick, Coventry, CV4 7AL, United Kingdom.}

\date{\today}

\begin{abstract}
A major challenge in the study of active systems is to harness their non-equilibrium dynamics into useful work. We address this by showing how to design colloids with controllable spontaneous propulsion or rotation when immersed in active nematics. This is illustrated for discs with tilted anchoring and chiral cogs, for which we determine the nematic director through conformal mappings. Our analysis identifies two regimes of behaviour for chiral cogs: orientation-dependent handedness and persistent active rotation. Finally, we provide design principles for active nematic colloids to achieve desired rotational dynamics.
\end{abstract}
\maketitle

\section{Introduction}
\label{sec:intro}

A broad class of both biological and synthetic materials are maintained out of equilibrium by motile constituents which exhibit orientational ordering \cite{ramaswamy2010mechanics,marchetti2013hydrodynamics,doostmohammadi2018active}, including cell monolayers~\cite{duclos2017topological}, tissues~\cite{saw2017topological}, bacteria in liquid crystalline environments~\cite{zhou2014living} and bacterial suspensions~\cite{wensink2012meso}, and synthetic suspensions of microtubules~\cite{sanchez2012spontaneous}. Such materials may be modelled as active liquid crystals, of which active nematics \cite{doostmohammadi2018active} have garnered the most attention, although consideration has also been given to smectic \cite{adhyapak2013live,chen_toner2013}, cholesteric \cite{whitfield2017hydrodynamic,kole2021layered} and hexactic \cite{maitra2020} phases. In active nematics the fundamental excitations are topological defects \cite{duclos2020topological,vcopar2019topology}, which acquire self-propulsive \cite{narayan2007,giomi2013defect,giomi2014defect,binysh2020three} and self-orienting \cite{houston2022defect} dynamics. They have also been shown to be foci for biological functionality, for example in epithelial cell apoptosis \cite{saw2017topological}, morphogenesis \cite{maroudas-sacks2021,guillamat2022integer} and the formation of bacterial colonies \cite{doostmohammadi2016defect,dell2018growing,basaran2022large} and biofilms \cite{yaman2019emergence}. Consequently, much effort has gone into controlling active nematics through confinement \cite{shendruk2017dancing,norton2018insensitivity,opathalage2019self}, external fields \cite{guillamat2017taming}, geometry and topology \cite{keber2014topology,ellis2018curvature}.

An ongoing challenge across all forms of active matter is to devise means of rectifying the non-equilibrium dynamics of active systems, allowing for the extraction of useful work and the development of active micromachines \cite{zhang2021autonomous,lv2022nano}. Such a harnessing of active dynamics is strikingly typified by bacterial ratchets \cite{di2010bacterial,sokolov2010swimming} -- the spontaneous and persistent rotation of chiral cogs immersed in a bath of bacteria, a phenomenon that is precluded in equilibrium systems. These experiments were performed with an isotropic, unordered collection of bacteria. In light of the many active systems which exhibit nematic ordering, it is important to establish whether colloids are still able to produce coherent dynamics from microscale activity and in particular if the same active ratchet effects can occur. 

There is a long and successful history of using colloids to modify the behaviour of passive nematics \cite{poulin1997novel,stark2001physics,muvsevivc2017liquid}, allowing the formation of metamaterials \cite{muvsevivc2006two,ravnik2013confined}, the control of topological defects \cite{ravnik2007,tkalec2011reconfigurable,muvsevivc2017liquid} and the tuning of colloidal geometry to induce certain elastic distortions \cite{lapointe2009shape,lapointe2013star}. However, the understanding of how to use colloidal inclusions to induce prescribed dynamics is in comparative infancy. There has been work on the dynamics of colloids in scalar active matter \cite{baek2018generic} as well as active droplets in nematic environments \cite{guillamat2018,hardouin2019,rajabi2020directional} and various realisations of propulsive \cite{loewe2022passive,yao2022topological} and rotational \cite{thampi2016activeb,ray2023} colloids in active nematics, but a theoretical understanding of how to design a colloid to produce a desired dynamical response is lacking. Here we build upon our recent `active nematic multipole' framework for determining active responses to eluicidate the connection between colloidal design and the induced dynamical response, focusing on discs with tilted anchoring and chiral cogs. Of particular relevance as a point of comparison for our work is the recent demonstration of persistent rotation for chiral colloids in active nematics \cite{ray2023}.

Our main findings are that spontaneous propulsion and rotation are generic properties of colloids in active nematics and can be tuned by either anchoring conditions or colloidal geometry. We illustrate the role of anchoring through discs with a uniformly tilted boundary condition for the director, showing that this tilt angle naturally determines the direction of propulsion or the magnitude and sign of rotation. The effect of colloidal geometry is demonstrated using chiral cogs, for which we find two regimes of active response - orientation-dependent chirality for cogs with few teeth and persistent rotation for those with more. We also provide optimal design principles for generating active rotation via cogs.

The remainder of this paper is organised as follows. In Section \ref{sec:active_multipoles} we present a summary of our previous work on active nematic multipoles \cite{houston2023active}, showing how dipole distortions produce net active forces and propulsion while quadrupoles lead to torques and rotation. In Section \ref{sec:colloidal_discs} we determine analytically the energy minimising director in the presence of colloidal discs and show how the multipole structure and hence active response can be determined by the director anchoring. Additionally, this section provides an intuition for our subsequent results concerning chiral cogs, since, in the limit of many teeth, their behaviour converges onto the `classical limit' of discs with tilted anchoring. Section \ref{sec:Cog Polygons}, in which we study cog colloids, forms the main part of the paper. In it we first construct the conformal maps and principles for the lowest energy director boundary conditions that we need to determine the induced director distortion. Then we once again perform a multipole expansion of the director to infer the active response of the cogs, before discussing how this is augmented by the number of teeth, their angle and boundary conditions. Section \ref{sec:discussion} contains a summary and discussion.

\section{Active Nematic Multipoles, Forces and Torques}
\label{sec:active_multipoles}

We summarise briefly the description of two-dimensional active nematic multipole states and their connection to active forces and torques~\cite{houston2023active}. In a minimal analytical approximation the director field is taken to have its equilibrium form and the effect of activity is captured through the flows induced by the usual active stress, $\boldsymbol{\sigma}^{\textrm{a}} = - \zeta {\bf nn}$, for that equilibrium director~\eqref{eq:MultipoleExpansion}. 
We consider director fields that can be linearised around a (locally) uniformly aligned state, ${\bf n} = {\bf e}_y + \delta n \,{\bf e}_x$. Taking a one-elastic constant Frank free energy, the equilibrium condition on the director field is that $\delta n$ should be harmonic, $\nabla^2 \delta n = 0$. We write the general solution as a multipole expansion 
\begin{equation}
    \delta n=A\frac{\ln(r/R)}{\ln(a/R)}+\frac{1}{2}\sum_{l=0}^{\infty}a^l\left(c_l\partial^l_z\ln r+\bar{c}_l\partial^l_{\bar{z}}\ln r\right),
    \label{eq:MultipoleExpansion}
\end{equation}
where $a$ is a characteristic length scale associated with the `core region', and $\partial_z=1/2(\partial_x-\text{i}\partial_y)$, $\partial_{\bar{z}}=1/2(\partial_x+\text{i}\partial_y)$ are the usual derivatives in the complex variable $z=x+\mathrm{i}y$. The first term in~\eqref{eq:MultipoleExpansion} is a monopole distortion whose strength is set by a large lengthscale $R$. At each order there are two distinct multipoles. For example the two dipoles may be written as $x/r^2$ and $y/r^2$, or equivalently as the real and imaginary parts of $\partial_z \ln r$. The complex coefficient $c_l$ specifies the weightings of the two multipoles at order $l$, such that $\frak{R}\left\lbrace c_l\right\rbrace$ provides the weighting of $\frak{R}\left\lbrace \partial_{\bar{z}}^l \ln r \right\rbrace$ and similarly for the imaginary part. 

Linearising in the director deformation $\delta n$, the active flows are given by the continuity and Stokes equations 
\begin{gather}
\nabla \cdot {\bf u} = 0 , \label{eq:continuity} \\
- \nabla p + \mu \nabla^2 {\bf u} = \zeta \bigl[ {\bf e}_y \partial_x \delta n + {\bf e}_x\partial_y \delta n \bigr] , \label{eq:active_Stokes_linearised}
\end{gather}
where ${\bf u}$ is the fluid velocity, $p$ the pressure and $\mu$ an isotropic viscosity. The activity is extensile when $\zeta$ is positive and contractile when $\zeta$ is negative. 
In the linearised Stokes equation~\eqref{eq:active_Stokes_linearised} the activity has the structure of a multipole expansion of derivatives on the monopole source $\ln r$, obtained from the expansion of $\delta n$~\eqref{eq:MultipoleExpansion}, and the general solution can therefore be given by acting with the same derivatives of the fundamental response to a monopole~\cite{houston2023active}. In two dimensions this fundamental active response is~\cite{houston2023active} 
\begin{align}
\begin{split}
    \Tilde{{\bf u}} & = \frac{\zeta\phi_0}{8\mu\ln(a/R)} \biggl[ \frac{x^2-y^2}{r^2} \bigl(-y \,\mathbf{e}_x + x \,\mathbf{e}_y \bigr) \\
    & \qquad + 2\ln \frac{r}{R} \,\bigl( y \,\mathbf{e}_x + x \,\mathbf{e}_y \bigr) \biggr] ,  
\end{split} \label{eq:active_flow_2D} \\
    \Tilde{p} & = -\frac{\zeta \phi_0}{\ln(a/R)} \frac{xy}{r^2} ,
    \label{eq:active_pressure_2D}
\end{align}
where $\phi_0$ is the angle of rotation of the director on the surface $r=a$. 
This minimal analytic approach has been found to be effective in characterising the active flows around defects in two dimensions~\cite{giomi2014defect,angheluta2021role}, defect loops in three dimensions~\cite{binysh2020three,houston2022defect}, in active turbulence~\cite{alert2020universal}, and on curved surfaces~\cite{khoromskaia2017vortex}. It may be considered to correspond to a limit of weak activity, however, even when there are defects, their structure may remain close to that in equilibrium. 

We can gain general insight into the active response of the nematic by considering the net contributions of the active stresses to the force and torque when integrated over a circle of radius $r$. This active force and torque are 
\begin{gather}
    \begin{split}
        \int \zeta \,\mathbf{n}\mathbf{n} \cdot \mathbf{e}_r \,r\text{d}\theta &\approx \int \zeta \left\lbrace \frac{y\delta n}{r} \mathbf{e}_x + \frac{x\delta n}{r} \mathbf{e}_y \right\rbrace r\text{d}\theta \\
        &=\frac{\zeta a\pi}{2}\left[\frak{I}\left\lbrace c_1\right\rbrace\mathbf{e}_x+\frak{R}\left\lbrace c_1\right\rbrace\mathbf{e}_y\right],
    \end{split} \label{eq:ActiveForce} \\
    \begin{split}
        \int \mathbf{x} \times \zeta \mathbf{n}\mathbf{n} \cdot \mathbf{e}_r \,r\text{d}\theta &\approx \int \zeta \frac{(x^2-y^2)\delta n}{r} \,r\text{d}\theta \\
        &=-\frac{\zeta a^2\pi}{2}\frak{R}\left\lbrace c_2\right\rbrace ,
    \end{split}
    \label{eq:ActiveTorque}
\end{gather}
and are determined, respectively, by the dipole and quadrupole coefficients $c_1$ and $c_2$. 
We see that in two dimensions both dipoles will self-propel if free to move and that there is a single chiral quadrupole which produces a rotational response. Comparing with \eqref{eq:MultipoleExpansion} we see that motion along $\mathbf{e}_x$ and $\mathbf{e}_y$ is dictated by the imaginary and real parts of $c_1$ respectively, such that the magnitude and phase of $c_1$ determine the speed and direction of motion. Rotational effects are governed by $\frak{R}\left\lbrace c_2\right\rbrace$, which we denote by $C_{\circlearrowleft}$, where the arrow indicates the sense of rotation that results due to a positive chiral quadrupole coefficient in an extensile system.
Consequently, much of the remainder of this paper is dedicated to means of controlling this chiral quadrupole coefficient, considering first discs with tilted boundary conditions and then chiral cogs with normal anchoring. Since the discs with tilted anchoring are a high-side-number limit of the chiral cogs, understanding their behaviour will provide intuitive principles for the more complicated cog dynamics. 

\section{Disc Colloids With Varying Anchoring Conditions}
\label{sec:colloidal_discs}

In this section we determine the energy-minimising director in the presence of a colloid. We can then read off the relevant multipole coefficients and infer the active response from the results of Section \ref{sec:active_multipoles}. With the aim of understanding both the dipole propulsive and quadrupole rotational active response we consider a colloid first with a single companion bulk defect and then with a pair of antipodal defects.

The director is specified by an angle $\phi$, which in a one-elastic-constant approximation must be a harmonic function. To maintain consistency with our far-field description $\mathbf{n}=\mathbf{e}_y+\delta n\mathbf{e}_x$ we define the director angle such that $\mathbf{n}=-\sin\phi \,\mathbf{e}_x + \cos\phi \,\mathbf{e}_y$, with $\phi$ vanishing asymptotically. In the far-field region the correspondence between the exact solution and the multipole expansion of \eqref{eq:MultipoleExpansion} is therefore given by $\delta n=-\phi$.

\subsection{Spontaneous Motion}
\label{subsec:disc_motion}

We begin by considering a colloidal disc of radius $a$ accompanied by a single bulk defect whose charge compensates the topological index of its boundary condition. We take the boundary condition to be one of uniform tilt from normal anchoring by an angle $\alpha$. This boundary condition has index $+1$ and so the colloid is accompanied by a $-1$ defect which we take to have a position described by the complex number $z_d=r_d e^{\text{i}\theta_d}$, with $\theta_d$ measured anticlockwise from the far-field direction, $\mathbf{e}_y$. The director angle may be written as $\phi=\frak{I}\left\lbrace\ln\Phi\right\rbrace$, with $\Phi$ a meromorphic function \cite{alexander2018topology}. In particular, by using the image system appropriate to a disc, one finds that the solution commensurate with the colloidal boundary condition we have described is
\begin{equation}
    \phi=\frak{I}\left\lbrace\ln\left[\frac{z^2}{(z-z_d)(z-\frac{a^2}{\bar{z}_d})}\right]\right\rbrace+(\theta_d+\alpha)\frac{\ln(\sqrt{z\bar{z}}/R)}{\ln(a/R)}.
    \label{eq:DipoleDirectorAngle}
\end{equation}
This solution has one complex free parameter $z_d$, to be determined by minimisation of the Frank free energy as described in the Appendix and considered previously in the context of smectic C films \cite{pettey1998topological}. 

\begin{figure*}
    \centering
    \includegraphics[width=1\linewidth, trim = 0 0 0 0, clip, angle = 0, origin = c]{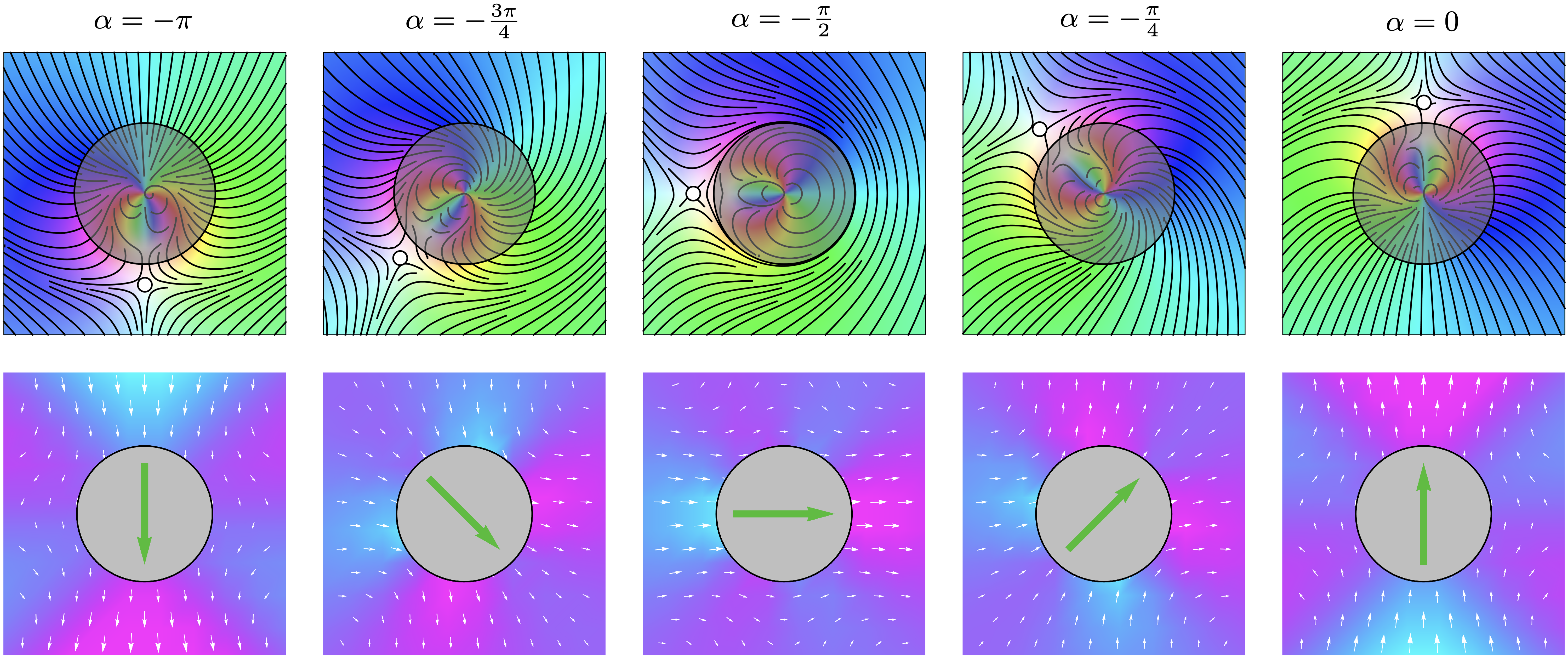}
    \caption{Analytical director configurations and active flows for a disc with a single bulk defect and varying anchoring angle $\alpha$. In the top row the integral director curves are shown in black upon a phase field of the director angle. The angular position of the bulk defect varies commensurately with the anchoring angle, but with opposite sign. Below, the far-field active flow is shown superposed on the pressure. The green arrows indicate the direction of self-propulsion for a system with extensile active stresses, which is always towards the bulk defect.
    }
    \label{fig:DiscDipole}
\end{figure*}

Expanding the logarithm in \eqref{eq:DipoleDirectorAngle} we find that the dipole coefficient is given by
\begin{equation}
    c_1=-2\frac{r_d^2+a^2}{ar_d}\text{e}^{\text{i}\theta_d},
\end{equation}
such that as expected, the dipole strength is set by the distance of the bulk $-1$ defect from the virtual $+2$ defect at the centre of the colloid, while the character of the dipole is determined by the orientation of this separation vector with respect to the far field. While all functions of the form of \eqref{eq:DipoleDirectorAngle} provide a minimum energy configuration for a given bulk defect location, there is still the question of which defect position provides the global energy minimiser for a specific colloidal anchoring condition. As determined in the Appendix, the Frank free energy is minimised by $\theta_d=-\alpha$ and $r_d=\sqrt{2}a$ \cite{pettey1998topological}, such that
\begin{equation}
    c_1=-3\sqrt{2}\text{e}^{-\text{i}\alpha}.
\end{equation}
As illustrated in Figure \ref{fig:DiscDipole}, this has the consequence that the direction of colloidal self-propulsion is directly determined by the anchoring condition. Note that, although the boundary condition is invariant under changing $\alpha$ by $\pi$, the dipole distortion is not. There are therefore two equal energy defect locations for any boundary condition, whose dipole coefficients differ by a sign. This is shown in Figure \ref{fig:DiscDipole} for normal anchoring, for which the bulk defect may sit below or above the colloid, with the motion of the colloid being towards the companion defect in an extensional system. However, between these two configurations the position vector of the bulk defect and the self-propulsion direction rotate in opposite senses so that generically the colloid does not propel towards the defect. This counterrotation of the propulsion direction compared to the bulk defect location has been observed in the context of Janus colloids \cite{loewe2022passive} and can be understood by making an association between dipole distortions and pairs of $\pm 1/2$ defects \cite{houston2023active}. In contractile systems the colloidal motion is reversed, such that it is towards the companion defect for tangential anchoring and away from it for normal anchoring.

\subsection{Spontaneous Rotation}
\label{subsec:disc_rotation}

\begin{figure*}
    \centering
    \includegraphics[width=1\linewidth, trim = 0 0 0 0, clip, angle = 0, origin = c]{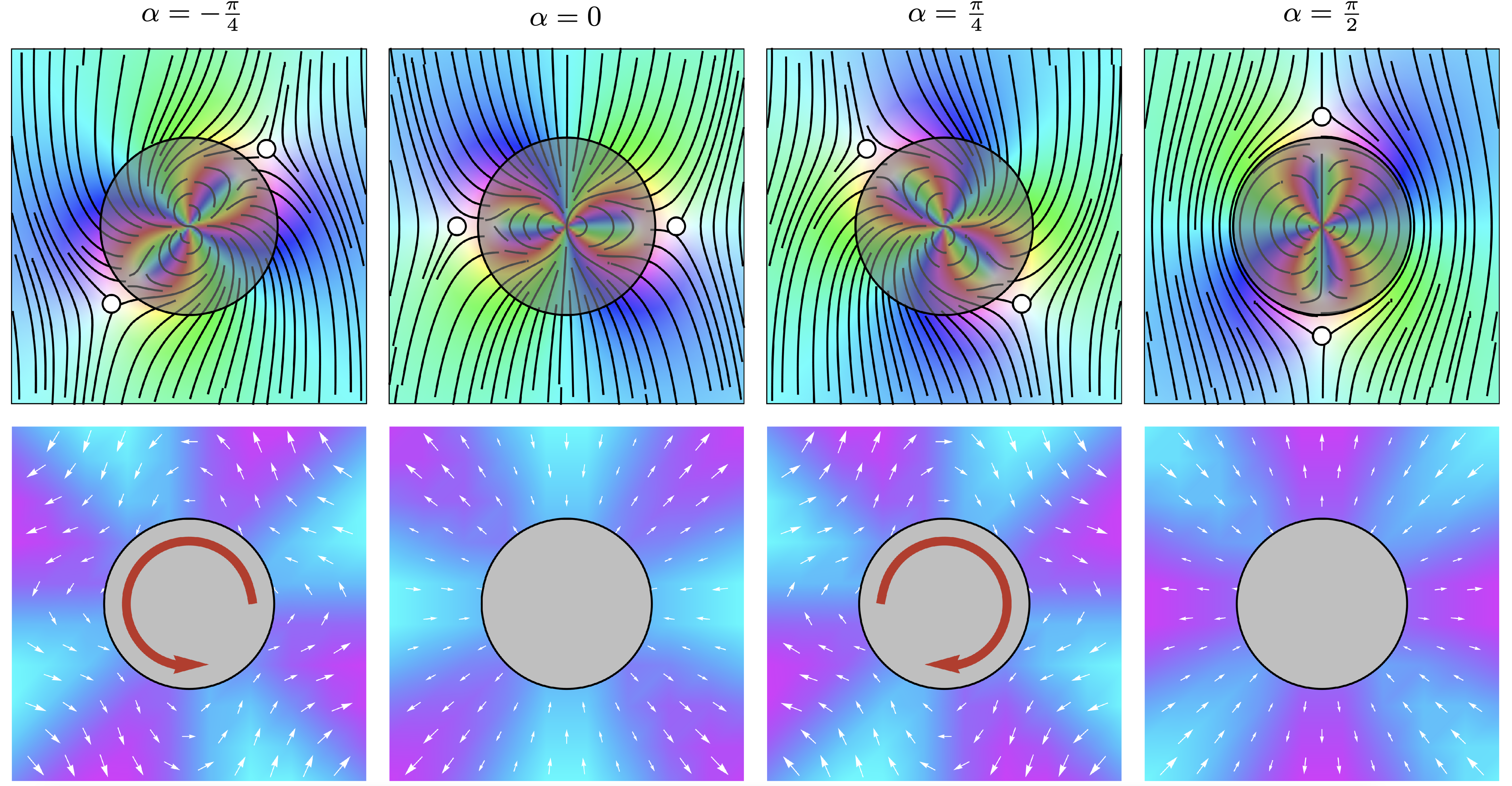}
    \caption{Analytical director configurations and active flows for a disc with two bulk defects and varying anchoring angle $\alpha$. The top row shows the integral director curves superposed on a phase field of the director angle. The angular positions of the bulk defects vary commensurately and in opposition with the anchoring angle. Below, the far-field active flow is shown upon the pressure. The red arrows indicate the direction of self-rotation when the active stresses are extensile.
    }
    \label{fig:DiscQuadrupole}
\end{figure*}

In the above case the dipole term always dominates. For a situation where the quadrupole is dominant we consider a disc with two $-1/2$ bulk defects. Taking these to be at positions $z_{d1}=r_{d1}\text{e}^{\text{i}\theta_{d1}}$ and $z_{d2}=r_{d2}\text{e}^{\text{i}\theta_{d2}}$ the appropriate harmonic functions have the form
\begin{equation}
    \begin{split}
        \phi&=\frak{I}\left\lbrace\ln\left[\frac{z^2}{\sqrt{(z-z_{d1})(z-a^2/\bar{z}_{d1})(z-z_{d2})(z-a^2/\bar{z}_{d2})}}\right]\right\rbrace\\
        &+\left(\frac{\theta_{d1}+\theta_{d2}}{2}+\alpha\right)\frac{\ln(\sqrt{z\bar{z}}/R)}{\ln(a/R)}.
    \label{eq:QuadrupoleDirectorAngle}
    \end{split}
\end{equation}
The determination of the global energy minimiser can again be found in prior work \cite{fukuda2001director,tasinkevych2002colloidal}, although we provide the calculation in the Appendix, with the result being that the defects sit at antipodal points with angles $-\alpha\pm\pi/2$ and with a common radial displacement of $(7/3)^{1/4}a$. With this in place, expansion of \eqref{eq:QuadrupoleDirectorAngle} gives the quadrupole coefficient as
\begin{equation}
    c_2=-\frac{10}{\sqrt{21}}\text{i}\text{e}^{-2\text{i}\alpha}.
\end{equation}

As before, the character of the nematic multipole is determined by the director anchoring condition. That the quadrupole varies with $\alpha$ at twice the rate of the dipole is a direct consequence of its inversion symmetry, indeed for an order $l$ multipole $c_l\sim \text{e}^{-\text{i}l\alpha}$. Consequently the net active torque varies with anchoring angle as $\sin2\alpha$. The active response is shown in Figure \ref{fig:DiscQuadrupole}. The chiral quadrupole coefficient $C_{\circlearrowleft}$ may be inferred from the defect configuration as follows. When there is a vertical mirror line $C_{\circlearrowleft}$ is zero, when they are placed upper right and lower left $C_{\circlearrowleft}$ is positive and when positioned conversely it is negative. This can also be discerned from the director orientation, represented by the colour map. For a purely achiral distortion the colour map forms a checkerboard pattern, while when $C_{\circlearrowleft}$ is non-zero the colours form a cross, with the colours inverted when the sign of $C_{\circlearrowleft}$ changes. The active flow is purely radial for normal and tangential anchoring, while the angular flow component is maximal for anchoring angles of $\pm\pi/4$. Variation of the anchoring condition therefore allows both the sign and magnitude of the colloidal self-rotation to be determined. 

\section{Cog Colloids}
\label{sec:Cog Polygons}

From chirality via anchoring conditions on a disc, we now turn our attention to controlling chiral active responses geometrically using chiral cogs. In particular we seek to understand how active ratchet behaviour such as is observed in bacterial baths \cite{di2010bacterial,sokolov2010swimming} might be generated in nematically ordered systems, as recently demonstrated experimentally \cite{ray2023}. In Section \ref{sec:active_multipoles} we identified the chiral quadrupole as being responsible for the generation of a net active torque and so persistent actively-driven rotation will require a consistent sign of this multipole as the orientation of the colloid is varied.

\begin{figure}[t]
    \centering
    \includegraphics[width=1\linewidth, trim = 0 0 0 0, clip, angle = 0, origin = c]{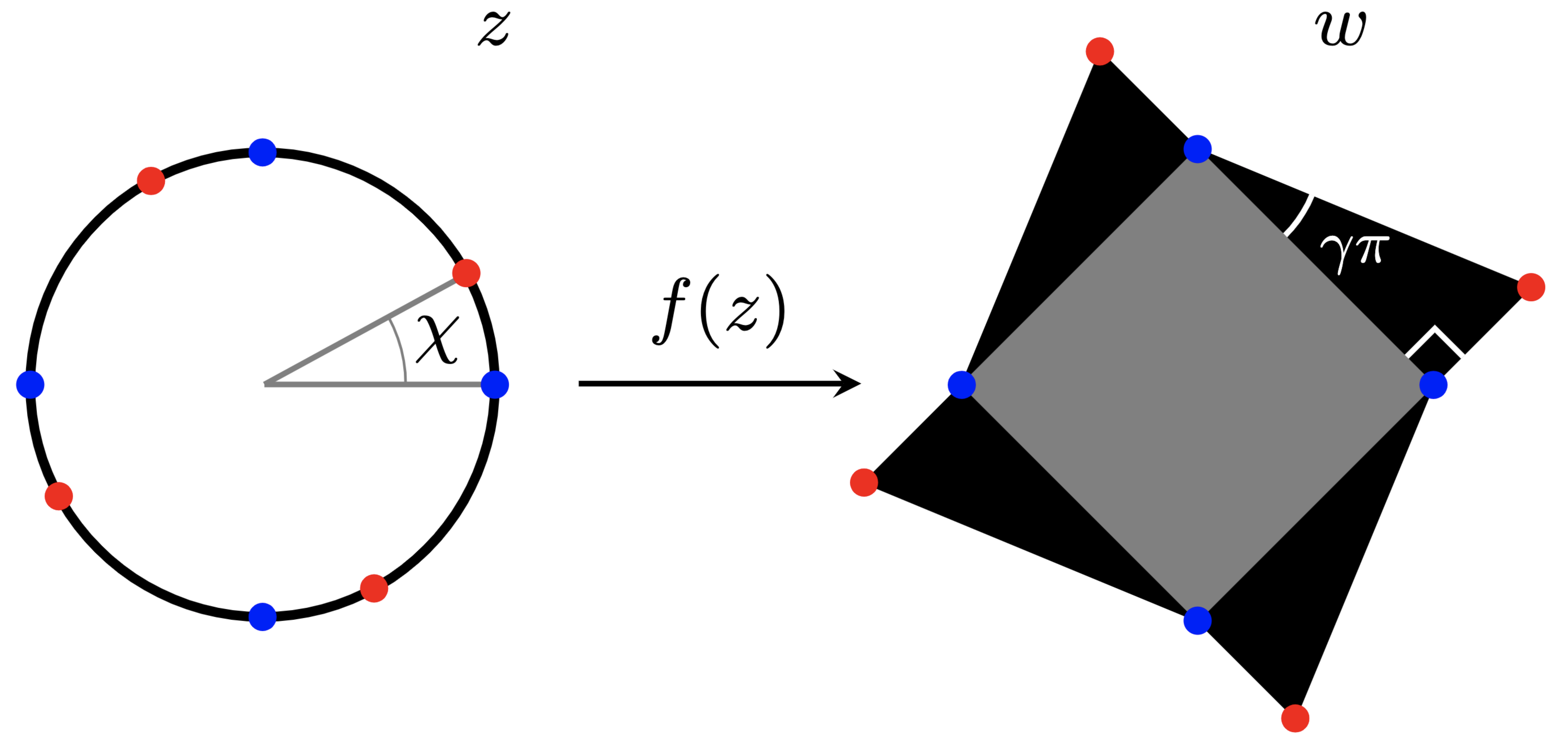}
    \caption{A schematic of a conformal mapping from the exterior of the disc to that of a cog. The blue and red points on the disc are mapped to the corresponding vertices of the cog, with the blue ones being fixed points of the map. The offset $\chi$ is dependent both on the number of cog teeth and their angle.
    }
    \label{fig:RepresentativePolygons}
\end{figure}

We find that the minimum energy director configuration changes discontinuously with the orientation of the cog whenever an edge passes through the far-field director orientation. In particular, this can lead to changes in the sign of the chiral quadrupole, a form of orientation-dependent chirality \cite{efrati2014orientation,dietler2020chirality}. In light of our preceeding results this presents an apparent impasse in our attempt to generate active ratchet effects. However, we show that as the number of cog teeth is increased this orientation-dependence fades away, the cogs map onto discs with tilted anchoring and persistent active rotation is recovered.

\subsection{Conformal mappings}
\label{subsec:Conformal Maps}

We construct a two-parameter family of chiral cogs from regular $n$-gons by adding a right-angled triangle with base angle $\gamma\pi$ to each edge, as illustrated in Figure \ref{fig:RepresentativePolygons} for a square-based cog with $\gamma=1/8$. The energy-minimising director angle is provided by a harmonic function which is uniquely determined a boundary condition corresponding to normal anchoring on the cog surface along with an asymptotic value provided by the far-field alignment. 

The solution for the cog is conveniently obtained via conformal mapping $f: \Omega \to \Omega^{\prime}$ from the exterior of a disc ($\Omega$) to the exterior of the polygon ($\Omega^{\prime}$). A holomorphic function $\phi(z)$ on $\Omega$ satisfying the appropriate piecewise boundary conditions on the disc will then provide the solution $\phi(f(z))$ for the polygon~\cite{ahlfors1953complex}. Provided we can construct the appropriate mapping, we can therefore separate the determination of the director field into two parts: finding conformal maps from the exterior of the disc onto the exterior of our colloid and solving the analogous Dirichlet problem for the director angle on the exterior of the disc.

To facilitate the construction of explicit mappings we use the framework of Schwarz-Christoffel transformations~\cite{ahlfors1979introduction,driscoll2002schwarz}. 
The relevant elements of the theory along with the derivation of the requisite conformal map are provided in the Appendix. Here we simply present the result that a conformal map $f: \Omega \to \Omega^{\prime}$ from the exterior of the disc ($\Omega$) to the exterior of a cog ($\Omega^{\prime}$) is given by
\begin{widetext}
\begin{equation}
    f(z)=\frac{\Gamma\left(\frac{1}{2}+\frac{1}{n}-\gamma\right)zF_1\left(-\frac{1}{n},\gamma+\frac{n-4}{2n},-\frac{1}{2}-\gamma,1-\frac{1}{n};\text{e}^{\text{i}n\psi}z^{-n},\text{e}^{\text{i}n(\psi+\chi)}z^{-n}\right)}{\Gamma\left(1-\frac{1}{n}\right)\Gamma\left(\frac{1}{2}+\frac{2}{n}-\gamma\right){}_2F_1\left(-\frac{1}{n},-\frac{1}{2}-\gamma,1-\frac{1}{n}-\frac{n-4}{2n}-\gamma,\text{e}^{\text{i}n\chi}\right)},
    \label{eq:SCCog}
\end{equation}
\end{widetext}
where the angle $\psi$ incorporates rotations of the cog. Here ${}_2F_1$ and $F_1$ denote respectively Gauss's hypergeometric function and the first Appell hypergeometric series of two variables \cite{NIST:DLMF}. This transformation is illustrated in Figure \ref{fig:RepresentativePolygons}, with the vertices of the cog and their corresponding prevertices on the disc highlighted. For the vertices of the base $n$-gon the prevertices sit at the $n^{\text{th}}$ roots of unity, those of the teeth tips being rigidly rotated by an angle $\chi$, dependent on both $n$ and $\gamma$.

\subsection{Director field and orientation-dependent handedness}
\label{subsec:Director field}

\begin{figure}[t]
    \centering
    \includegraphics[width=1\linewidth, trim = 0 0 0 0, clip, angle = 0, origin = c]{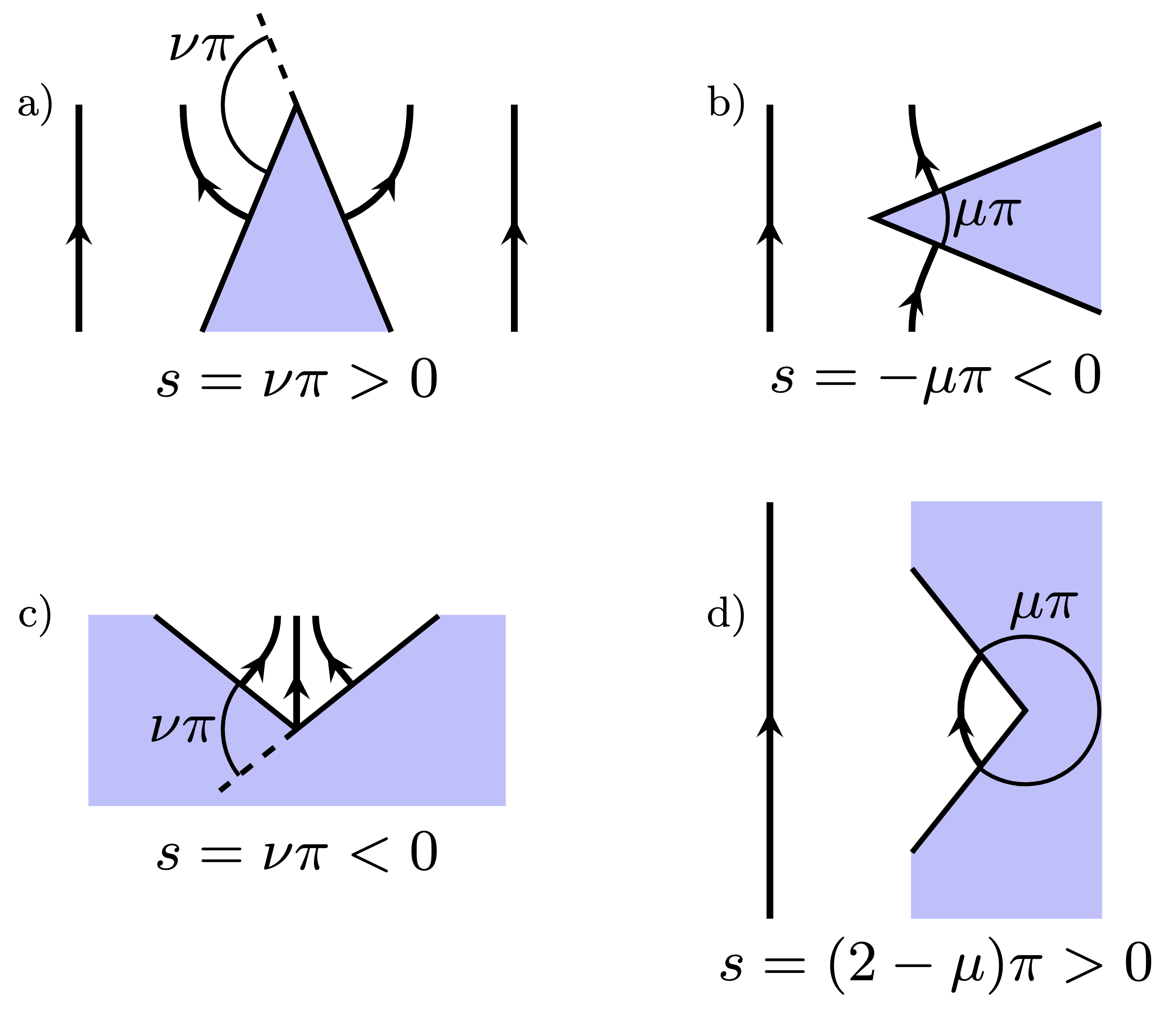}
    \caption[Local picture for director winding, $s$, at a vertex.]{Local picture for director winding at a vertex. Schematics of the director, decorated with an orientation to aid our argument, around a vertex of a colloid with normal anchoring and vertical far-field alignment. a) A convex vertex aligned along the far-field orientation induces positive director winding equal to the exterior angle. Rotating the vertex anticlockwise, the director orientation is reversed on the left-hand edge as it passes through the vertical. This rotation leaves the vertex pointing `into the flow' of the director winding, shown in b), with the induced winding now negative and equal in magnitude to the interior angle, having reduced by $\pi$. The story is similar for concave vertices, as shown in c) and d), except with a reversal of the signs of winding. When orientated along the far-field direction as in c) the winding is again equal to the exterior angle, this time negative. Upon rotation the director orientation along the right-hand edge flips, increasing the winding induced at the vertex by $\pi$.
    }
    \label{fig:VertexWinding}
\end{figure}

Before discussing the boundary conditions relevant to cogs it will be beneficial to consider the winding induced by an isolated vertex, as this will give insight into how the nematic distortion changes as a colloid rotates. It should be emphasised that we are treating the cog's rotation adiabatically, such that at every orientation the director adopts the minimal energy distortion. The first point to make is that the imposition of normal anchoring does not correspond to a unique Dirichlet boundary condition as at any point the director angle may be changed by a multiple of $\pi$ with the director still being normal to the colloid. However, the boundary condition appropriate for the minimal energy texture is that where the director angle $\phi\in\left[-\pi/2,\pi/2\right]$. Any other choice would correspond to additional topological defects on the boundary of the colloid and greater director distortion in connecting the value of $\phi$ on an edge to its asymptotic value.

Consequently, a vertex generically exists in one of two states, as illustrated in Figure \ref{fig:VertexWinding}, namely with the director on the two edges oriented in the same sense, either inwards or outwards, as in a) and c) or in opposing senses, as in b) and d). According to these circumstances we will have to rotate the director on one edge either clockwise or anticlockwise around the vertex to have it match that on the other side and hence the induced winding will be given by either the exterior angle $\nu\pi$ or interior angle $\mu\pi=(1-\nu)\pi$. The dividing line between these two regimes of behaviour is, in light of the bounds on $\phi$, when one of the edges of the vertex is vertical. As shown in Figure \ref{fig:VertexWinding}, when one of the edges of a vertex passes through the vertical an inversion event occurs in which the orientation on the edge is reversed and the director winding at the vertex changes by $\pm\pi$. Although we have chosen an orientation for the director to aid our discussion, all statements we make are of course independent of this choice.

This local description of the director winding at a vertex underpins the nature of the nematic distortions around all the colloids that we consider. As the orientation of the colloid is varied the distortion is rigidly rotated, with the addition of a monopole term to maintain the anchoring condition, along with director inversions whenever an edge is moved through the vertical so as to maintain the appropriate bounds on $\phi$. The redistribution of topological charge caused by these inversions provides a mechanism for orientation-dependent chirality \cite{efrati2014orientation,dietler2020chirality}, as we discuss further shortly.

\begin{figure*}
    \centering
    \includegraphics[width=1\linewidth, trim = 0 0 0 0, clip, angle = 0, origin = c]{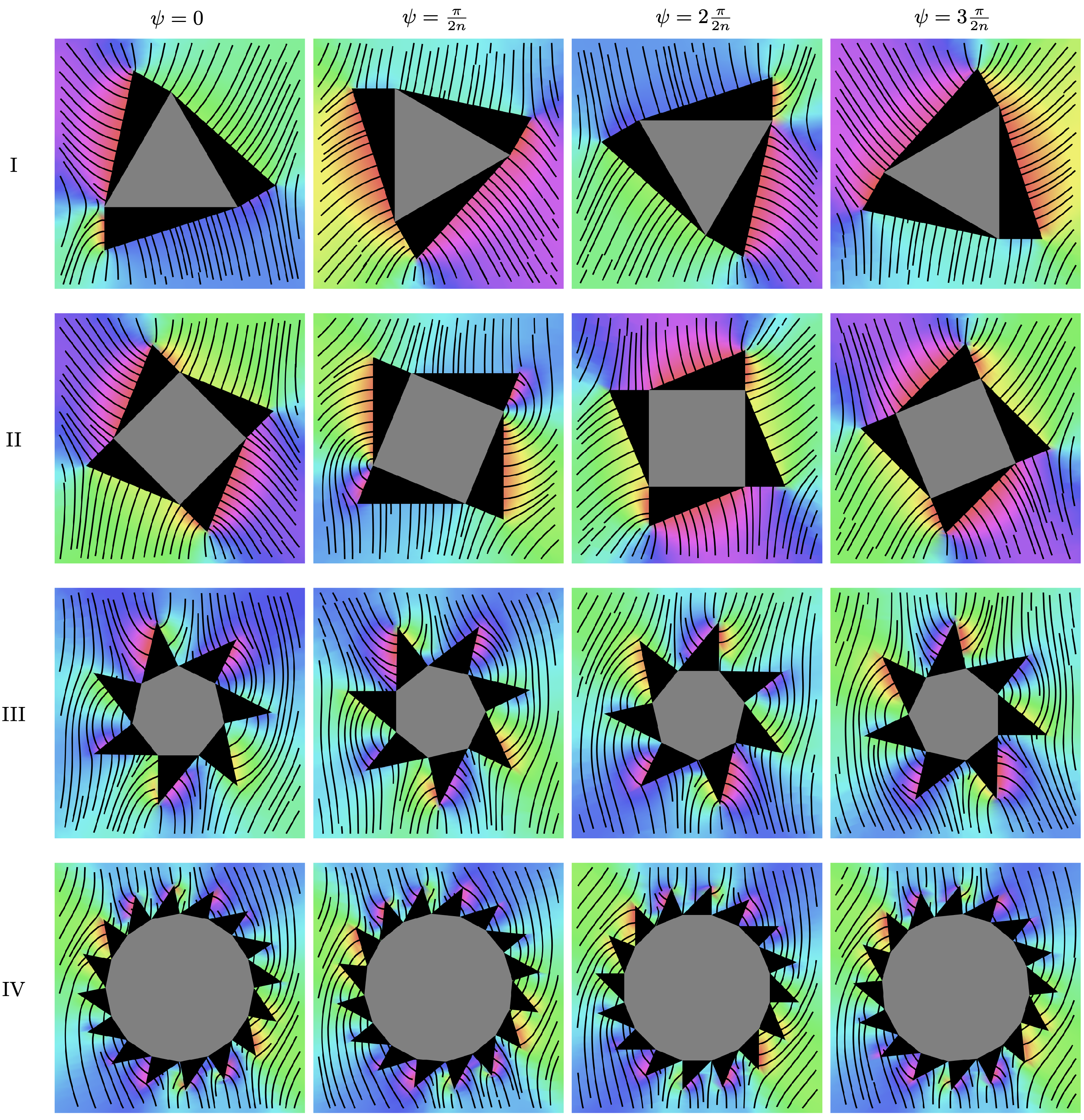}
    \caption{Analytical solutions for cog polygons in a two-dimensional nematic. The integral curves of the director are shown superposed on the phase field of its angle and the base regular polygon is shown inside in grey. Four equally-spaced orientations are shown for each cog, which sit at the locations of the polygonal icons in the phase space in Figure \ref{fig:CogsPhaseSpace}. Row I $(n=3,\gamma=0.1)$: the chiral quadrupole changes sign with rotation but is net negative, despite the teeth angle $\gamma$ being positive. Row II $(n=4,\gamma=0.125)$: the chiral quadrupole is positive on average but is not consistent in its sign. Row III $(n=7,\gamma=0.28$): the chiral quadrupole is consistently positive. This value of $\gamma$ gives the greatest average chirality for a cog with seven teeth, but there is still considerable variation as the cog rotates. Row IV $(n=16,\gamma=0.25)$: as the number of teeth increases the variation of the distortion due to rotation becomes smaller and smoother, the average chirality is maximised for $\gamma=0.25$ and this maximum value tends to $1/2$.
    }
    \label{fig:CogsAnalyticFurther}
\end{figure*}

With this understanding in place the boundary condition for a specific cog is not particularly illuminating and so rather than presenting it for each cog that we consider we give the algorithmic process for its construction. Taking the boundary condition for the base regular polygon as a starting point, the addition of the cog teeth is captured by a square wave alternating between the values $\gamma\pi$ and $-\pi/2$. This will typically result in a boundary condition where the director angle $\phi$ lies outside $[-\pi/2,\pi/2]$ in places and so the director orientation will need to be reversed along these segments to bring $\phi$ back within this range. The precise form of the boundary condition depends on the value of $n$ modulo $4$, but using the case $n\equiv 0$ mod $4$ as a partial example, the first step gives, for $\psi=0$,
\begin{widetext}
\begin{equation}
    \begin{split}
        \phi(\theta)&=\sum_{m=1}^{\infty}\frac{(-1)^m}{m}\sin(2m\theta)+\sum_{m=1}^{\infty}\frac{2}{mn}\sin(nm\theta)+\gamma\pi-\frac{(1+2\gamma)n\chi}{4}+\\
        &\sum_{m=1}^{\infty}\frac{\left(\frac{1}{2}+\gamma\right)\left(1+(-1)^m\right)}{m}\left[\sin\left[\frac{nm}{2}\left(\theta-\frac{n\chi}{2}\right)\right]-\sin\left(\frac{nm}{2}\theta\right)\right].
    \end{split}
\end{equation}
\end{widetext}
Any requisite reversal of director orientation may be achieved by using that
\begin{equation}
    \frac{\theta_2-\theta_1}{2}+\sum_{m=1}^{\infty}\frac{\sin(m(\theta-\theta_1))-\sin(m(\theta-\theta_2))}{m}
    \label{eq:EdgeInversion}
\end{equation}
increases $\phi$ by $\pi$ between $\theta_1$ and $\theta_2$.

Pairing the boundary conditions with the conformal map in \eqref{eq:SCCog} produces the director solutions shown in Figure \ref{fig:CogsAnalyticFurther}. A striking feature of the director solutions for triangle- and square-based cogs shown in Figure \ref{fig:CogsAnalyticFurther} is the reversal in sign of the chiral quadrupole, $C_{\circlearrowleft}$, an instance of orientation-dependent chirality \cite{efrati2014orientation,dietler2020chirality}. This is the multipole associated with net active torques and so this change in its sign strikes at the heart of our ability to use cogs to generate persistent rotations in active nematics. We therefore shall devote a little attention to understanding this behaviour.

In this paper we are using the induced director distortion, in particular the chiral quadrupole, as a measure of chirality for the colloid. Assigning a chiral measure is in general fraught with difficulty. There is not a unique choice, but rather an infinite hierarchy of possible measures \cite{harris1999molecular}, with the potential for different measures to determine distinct shapes as the `most chiral' \cite{fowler2005quantification}. For us, although higher-order terms could be invoked, the chiral quadrupole coefficient is a natural choice of chirality measure becuase it determines the active rotational response. It can also be connected to previously used measures. The chirality pseudotensor \cite{efrati2014orientation} reduces in the case of two-dimensional nematics to $\partial_i\phi$ \cite{long2021geometry}. Of this, the lowest-order term without mirror symmetry is the chiral quadrupole.

Our observation of orientation-dependent handedness of the nematic distortion raises a second issue with associating a chiral measure to a system -- that of chiral connectedness \cite{weinberg2000chirality,efrati2014orientation}. This occurs when a system can be continuously deformed from a chiral state to its mirror image without ever passing through an achiral configuration. This necessitates the chiral measure, which takes opposing signs for the mirror images, to pass through zero as its value is continuously changed and hence falsely label an intermediate chiral state as achiral. It is not inevitable that we encounter the paradox of chiral connectedness, since the chiral quadrupole can vary discontinuously with cog orientation. Nonetheless, we do find cog orientations for which the chiral quadrupole vanishes. However, in our context this is not so clearly a flaw, as it has a direct physical meaning -- in such orientations a cog, despite being chiral, experiences no net active torque.

There are two factors behind the variation of the chiral mode: the rigid rotation of the distortion as the cog is rotated and inversions of the director orientation along certain edges. Both of these can be reduced by increasing the number of teeth, as we now demonstrate. With $n$ teeth the cog returns to its original state after a rotation of $2\pi/n$, so the smooth variation of the chiral mode can be made arbitrarily small. Sticking with the case $n\equiv0$ mod $4$ for concreteness there will generically be two inversion events, each affecting a pair of antipodal edges. The exact length of the arc of the preimage circle corresponding to each edge depends on $\chi$ but is at most $2\pi/n$ and so from \eqref{eq:EdgeInversion} we can see that the modification to the boundary conditions necessitated by such an inversion is bounded by

\begin{equation}
    \begin{split}
        \frac{2\pi}{n}&+\sum_{m=1}^{\infty}\frac{1+(-1)^m}{m}\left[\left(1-\cos\left(\frac{2m\pi}{n}\right)\right)\sin(m(\theta-\theta_1))\right.\\
        &\left.+\sin\left(\frac{2m\pi}{n}\right)\cos(m(\theta-\theta_1))\right],
    \end{split}
\end{equation}
with $\theta_1$ varying according to the location of the edges on which the inversion is occurring. The discontinuous change brought about by these inversions can therefore also be made arbitrarily small. In short, the distortions induced by a cog converge onto the rotationally invariant disc distortion, with anchoring angle $\gamma\pi$. The orientational dependence of the chiral quadrupole is a signature of discretisation which grows ever fainter with increasing side number.

This smoothing out of distortion variations means that the persistent rotation of cogs observed in bacterial ratchets \cite{di2010bacterial,sokolov2010swimming} is achievable in orientationally-ordered active matter. An explicit demonstration of the consistent chirality of the induced distortion is given in Figure \ref{fig:CogsAnalyticFurther} for a cog with 16 teeth. The correspondence with the disc limit is confirmed by noting that the negative winding is situated on the first teeth to be tilted through the vertical, that is at $2\pi k/n$ where $k$ is the smallest integer such that $\pi/n+\gamma\pi+2\pi k/n>\pi/2$. In the large $n$ limit we may take equality and find that the displacement of the negative winding from the equatorial points tends to $-\gamma\pi$, in accordance with the quadrupole distortion for anchoring angle of $-\gamma\pi$ on the disc, as established in Section \ref{sec:colloidal_discs}.

\begin{figure*}
    \centering
    \includegraphics[width=1\linewidth, trim = 0 0 0 0, clip, angle = 0, origin = c]{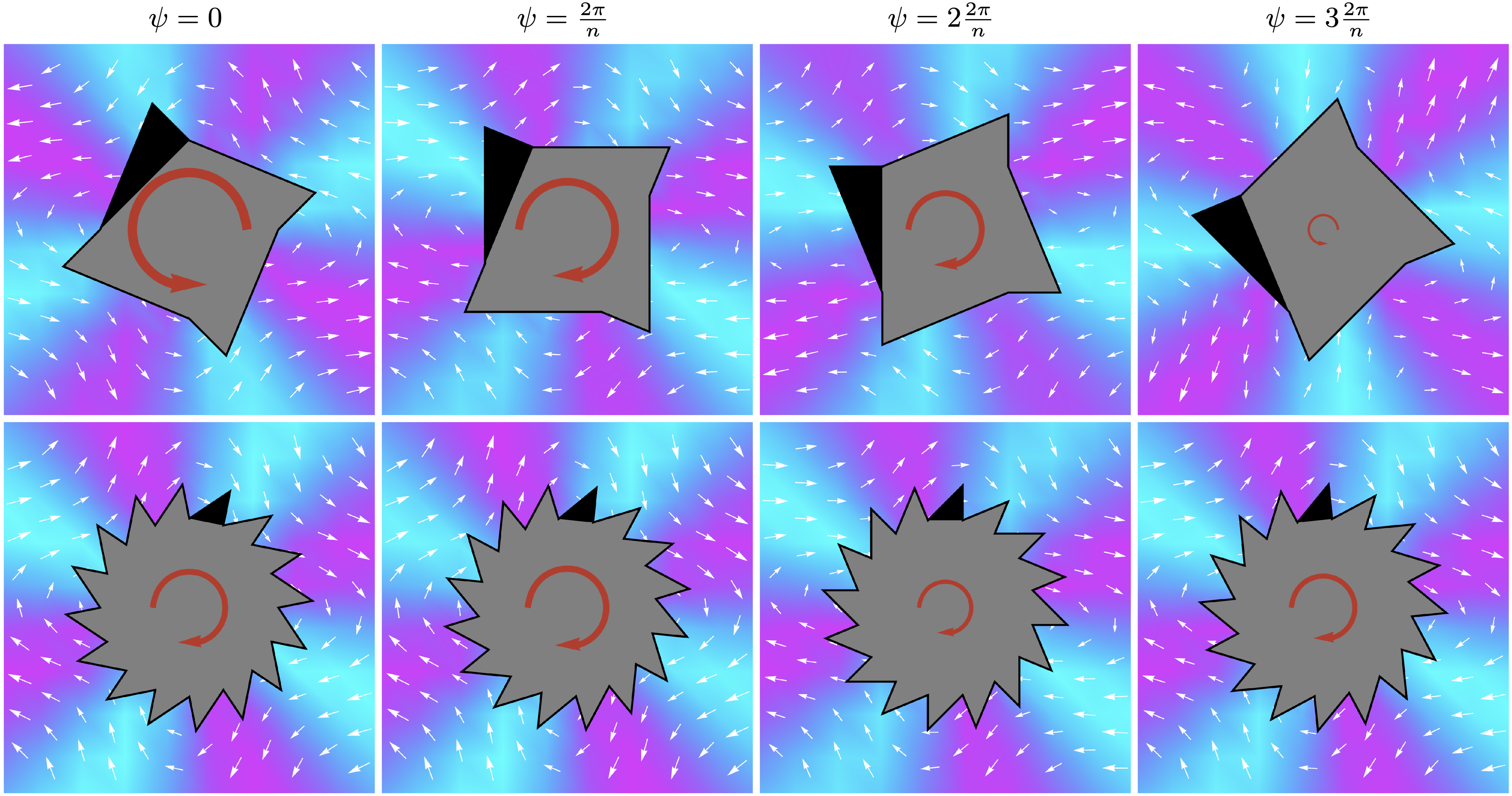}
    \caption{The far-field flows induced by cogs in extensile active nematics. Above, the square cog induces a flow whose rotational component changes sign, preventing consistent active self-rotation. Below, the active flow induced by the cog with sixteen teeth has a consistent sense of rotation, meaning that the cog exhibits persistent self-rotation. The background colour represents the pressure field, with the white arrows indicating the fluid flow. The red arrows show the direction and magnitude of the net active torque, while the black cog tooth is intended as a visual aid to make the changing orientation of the cog clearer.
    }
    \label{fig:CogsActiveFlow}
\end{figure*}

The consequences for active dynamics of this orientation-dependence of the chiral quadrupole are illustrated in Figure \ref{fig:CogsActiveFlow}. These far-field active flows are attained in the same fashion as those shown for discs in Section \ref{sec:colloidal_discs} -- from our exact solutions for the director around the cogs we read off the multipole coefficients to quadrupole order and can then determine the active response from the appropriate derivatives of \eqref{eq:active_flow_2D} and \eqref{eq:active_pressure_2D}. For the square-based cog, also shown in row II of Figure \ref{fig:CogsAnalyticFurther}, the discontinuous changes in the director field are of sufficient magnitude to change the sign of the chiral quadrupole and hence reverse the rotational component of the active flow, meaning the cog will not persistently rotate. By contrast, the consistent chiral quadrupole induced by the sixteen-toothed cog shown in row IV of Figure \ref{fig:CogsAnalyticFurther} results in a consistent sense of rotation in the active flow, allowing persistent active rotation of the cog. To be explicit, and as evidenced by the expression for the net active torque in \eqref{eq:ActiveTorque}, the direction of rotation is determined by the signs of both the chiral quadrupole and the activity. The flows and rotations shown in Figure \ref{fig:CogsActiveFlow} correspond to extensile activity and would be reversed in contractile systems.

\subsection{Active response}
\label{subsec:ActiveResponse}
\begin{figure}
    \centering
    \includegraphics[width=1\linewidth, trim = 0 0 0 0, clip, angle = 0, origin = c]{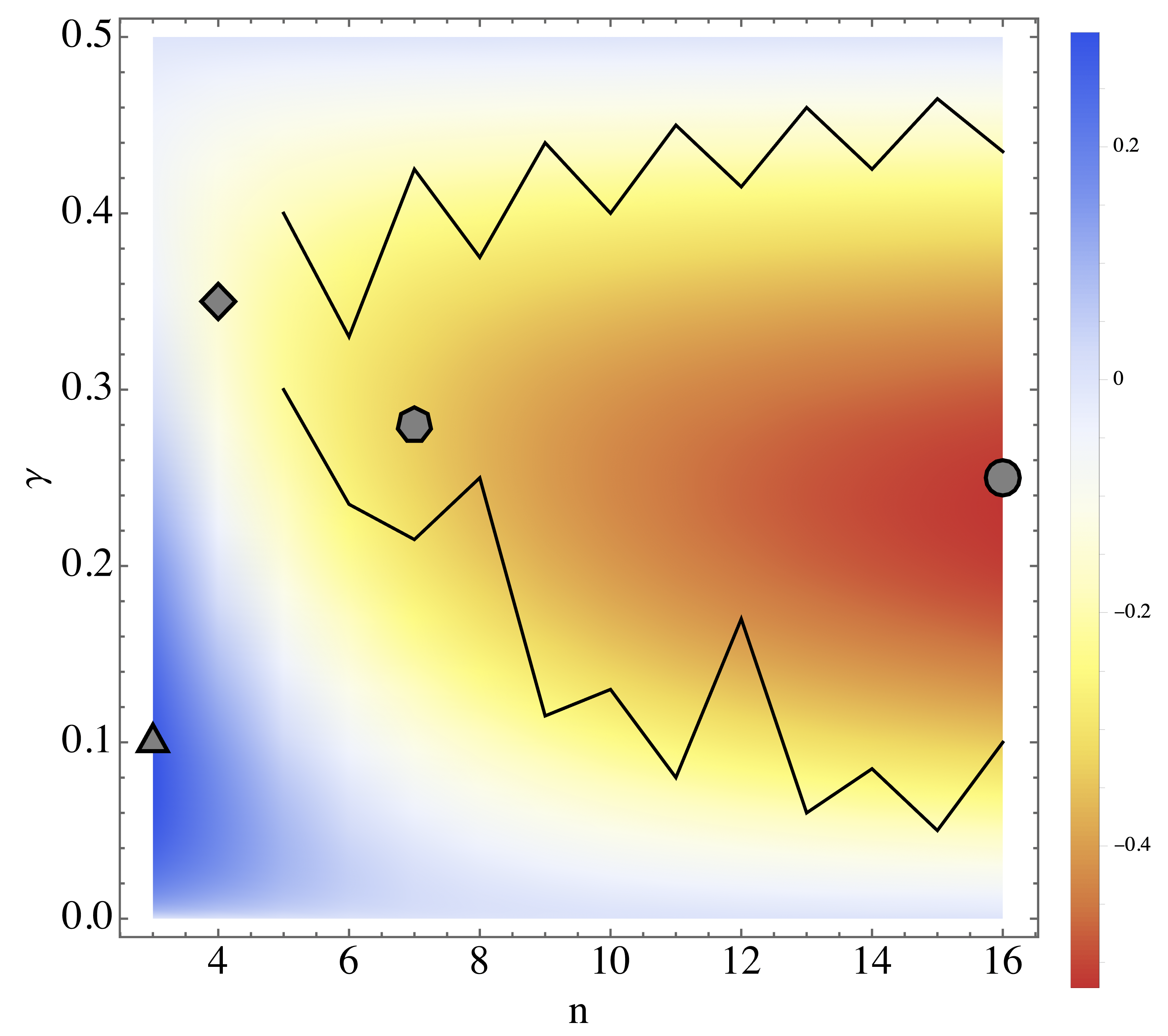}
        \caption[The phase space of the chirality of distortions induced by cogs in two-dimensional nematics.]{The phase space of the chirality of distortions induced by cogs in two-dimensional nematics. The temperature map represents the strength of the chiral quadrupole $C_{\circlearrowleft}$ when averaged over all orientations of the cog as a function of the number of cog teeth $n$ and the tooth angle $\gamma$. The two black lines form the boundaries of the region for which the chiral quadrupole has a consistent sign under rotation, corresponding to a consistent active torque in active nematics. The polygonal icons correspond to the locations in phase space of the cogs shown in Figure \ref{fig:CogsAnalyticFurther}.
    }
    \label{fig:CogsPhaseSpace}
\end{figure}

Having identified these two distinct modes of behaviour, let us investigate more closely the relationship between the chiral response of the active nematic and the parameters of the cog. Figure \ref{fig:CogsPhaseSpace} shows the phase space for the coefficient of the chiral quadrupole as a function of both the number of teeth $n$ and the tooth angle $\gamma\pi$. It was generated by constructing the boundary condition for a cog and then averaging the chiral quadrupole coefficient over a full rotation. This averaging was performed for 100 evenly-spaced values of $\gamma$ for each $n$ from $3$ to $16$. The distortions produced by the cogs which sit at the locations of the polygonal icons are shown in Figure \ref{fig:CogsAnalyticFurther}. Between the two black lines is the region for which the chiral quadrupole has a consistent sign under rotation, allowing for active nematic ratchets. There are several regions we can identify. When $n$ is small ($3$ or $4$) there is no value of $\gamma$ which will produce a consistent chiral response. The square icon denotes a typical example, shown in row II of Figure \ref{fig:CogsAnalyticFurther}; there is a modest net chiral quadrupole but its sign changes from the first two panels to the last two. The region in the bottom-left of Figure \ref{fig:CogsPhaseSpace} is even more extreme, as despite the angle of the cog teeth being such as to ostensibly induce a negative chiral quadrupole, the averaged coefficient is actually positive. This counter-intuitive state of affairs is illustrated by the triangle-based cog in row I of Figure \ref{fig:CogsAnalyticFurther}. The cog shown there is also convex, something that is possible only for triangle-based cogs with $\gamma<1/6$, but this is not necessary for the net chiral quadrupole to be positive. For $n\geq 5$ a band of $\gamma$ values develops for which the chiral response is of a consistent sign, with this band opening up with increasing $n$. The heptagon-based cog in row III of Figure \ref{fig:CogsAnalyticFurther} provides an example of the behaviour typical for intermediate values of $n$; the chiral quadrupole is of a consistent sign but there is still significant variation in the nematic distortion as the orientation of the cog is changed. Lastly, in the high $n$ limit we can understand the chiral response of the nematic to the cog by modelling it as a disc with anchoring angle $\gamma\pi$. Our discussion of discs in Section \ref{subsec:disc_rotation} predicts the dependence of the chiral quadrupole coefficient on $\gamma$ to be
\begin{equation}
    C_{\circlearrowleft}\sim-\sin(2\gamma\pi)
    \label{eq:CogChiralityHighN}
\end{equation}
and this is an increasingly good approximation for the behaviour in the high-$n$ region of parameter space. As demonstrated by the hexadecagon-based cog in row IV of Figure \ref{fig:CogsAnalyticFurther}, the variation of the distortion with cog orientation grows smaller and smoother with increasing $n$, the maximum value of the average chiral quadrupole tends to $1/2$ and the value of $\gamma$ for which this is achieved tends to $1/4$. As far as using cogs to generate active nematic ratchets the optimal design is therefore to have as many teeth as feasible, in order to maximise the smoothness of the net active torque and the range of $\gamma$ which can be utilised. The strength of the active torque can then be tuned using \eqref{eq:CogChiralityHighN}.

It should come as no surprise that the cogs which are `most chiral', at least in the sense of our definition, are those with intermediate values of $\gamma$. Both $\gamma=0$ and the limit $\gamma\to 1/2$ correspond to achiral shapes, the former regular polygons the latter tending to star-shaped polygons (although the $\gamma=1/2$ limit itself does not yield a bounded polygon, as the two edges of the cog teeth become parallel). The black lines in Figure \ref{fig:CogsPhaseSpace}, which bound the region of consistent chirality, are intriguingly non-monotonic and display sizeable jumps. The upper boundary exhibits an odd-even effect and is given by the largest fraction with denominator $n$ that is less than $1/2$. The lower boundary is less regular and seems to display a mod $4$ effect. Neither of these is particularly surprising, given that the compatibility of discrete boundaries with the symmetry of a quadrupole is inherently dependent on $n$ mod $4$ \cite{MyThesis}. We do not discuss the specifics of these threshold values any further since they are tied to the precise cog construction that we have employed in this paper and are thus of secondary interest compared to the qualitative phenomenology we have described, which applies to the chiral response of a nematic regardless of the exact geometry of the colloid.

\subsection{Role of anchoring conditions}
\label{subsec:Anchoring}

We conclude by discussing the effect of anchoring conditions on the active phenomenology of cogs. This is partly motivated by recent experimental realisations of cogs in active nematics \cite{ray2023}, which utilised tangential rather than normal anchoring. Once again the results we described for discs in Section \ref{sec:colloidal_discs} will prove insightful.

\begin{figure*}
    \centering
    \includegraphics[width=1\linewidth, trim = 0 0 0 0, clip, angle = 0, origin = c]{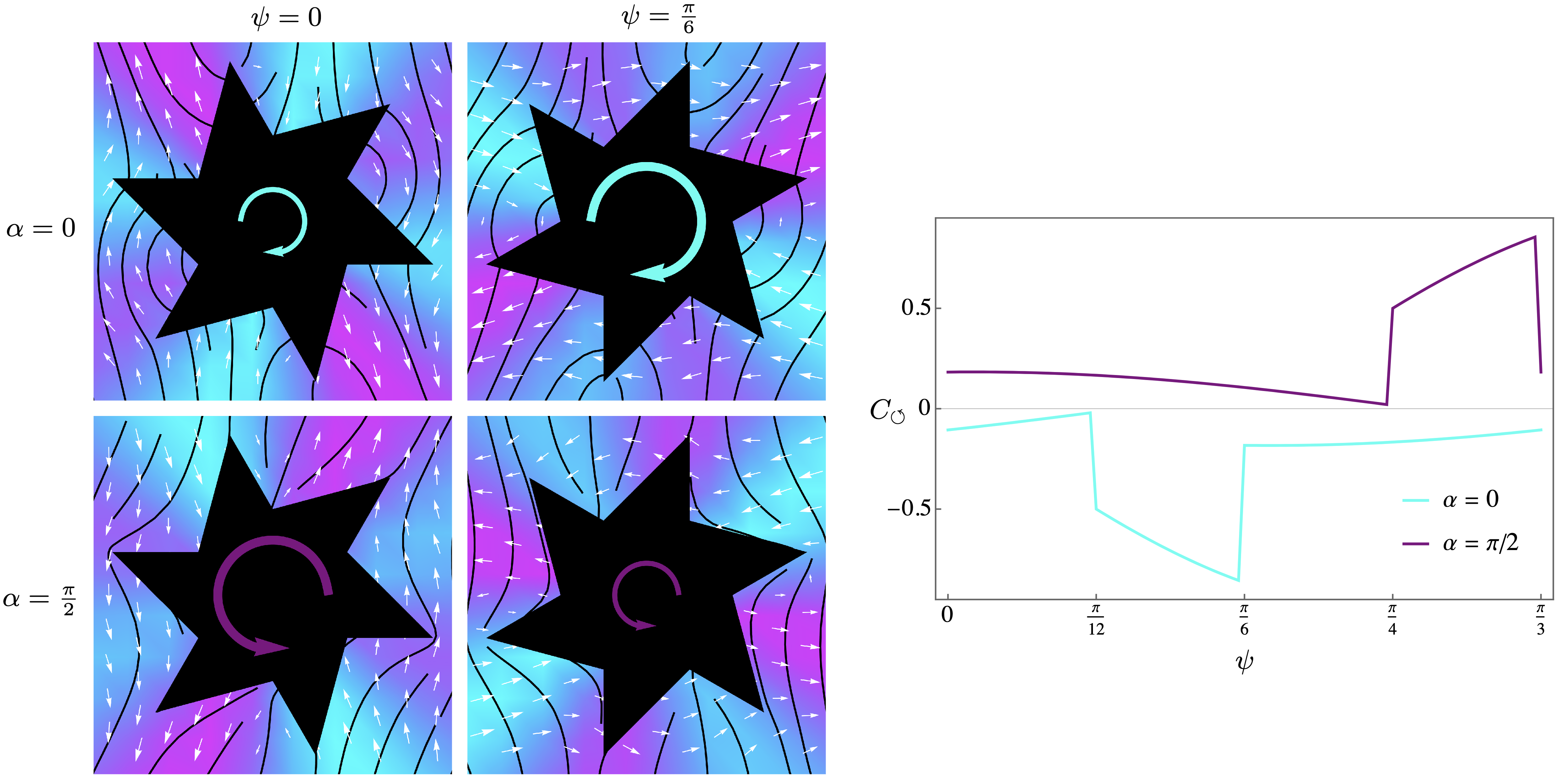}
    \caption{The effect of anchoring conditions on chiral active response. The panels on the left show the nematic director around a cog along with the induced active flow and pressure for different anchoring angles $\alpha$ and rotation angles $\psi$. The arrows indicate the direction and magnitude of active self-propulsion, dictated by the chiral quadrupole coefficient $C_{\circlearrowleft}$. The graph on the right plots this chiral quadrupole coefficient as a function of cog orientation for normal and tangential anchoring. 
    }
    \label{fig:Anchoring}
\end{figure*}

Having determined an energy-minimising director configuration for one anchoring condition we can find the minimal energy distortion for any other anchoring condition by applying the appropriate local rotation to the director. However, this transformation rotates the far-field direction, so to keep this fixed we must globally rotate the system in the opposite direction, changing the orientation of the cog. Denoting the anchoring and orientation angles by $\alpha$ and $\psi$ and recalling from Section \ref{sec:colloidal_discs} that for a director rotation by $\Delta$ the multipole coefficients obey $c_l\sim \text{e}^{-\text{i}l\Delta}$ we have the relation
\begin{equation}
    c_l(\alpha,\psi)=\text{e}^{-\text{i}l\Delta}c_l(\alpha-\Delta,\psi+\Delta).
    \label{eq:AnchoringConditionRelation}
\end{equation}

Of particular interest is the different effect normal or tangential anchoring has on the quadrupole coefficients. In this case \eqref{eq:AnchoringConditionRelation} reduces to
\begin{equation}
    c_2(\frac{\pi}{2},\psi)=-c_2(0,\psi+\frac{\pi}{2}).
\end{equation}
This relationship is illustrated in Figure \ref{fig:Anchoring}. The net active torque, proportional to $\frak{R}\left\lbrace c_2\right\rbrace$, as a function of orientation for normal anchoring is connected to that for tangential anchoring by reversing the sign and shifting by an angle of $\pi/2$. The cogs in Figure \ref{fig:Anchoring} have six-fold symmetry and so the angular shift appears as a displacement by $\pi/6$. Due to the reversal in sign of the chiral quadrupole, a cog with tangential anchoring will, all other properties being equal, rotate in the opposite direction to one with normal anchoring. The anticlockwise rotation illustrated in Figure \ref{fig:Anchoring} is in agreement with that seen in experiments \cite{ray2023}.

For any cog there is an optimal anchoring angle which maximises the strength of the net active torque. Following the principles laid out in Section \ref{sec:colloidal_discs} we can see that this optimised anchoring will be such that the director tilt angle and the cog tooth angle sum to approximately $\pi/4$. However, in cases where it is possible to control the anchoring to this extent it is simpler to just use disc colloids as there is no need to use colloidal geometry to induce chirality.

\section{Discussion}
\label{sec:discussion}

We have built upon a general framework for understanding the behaviour of distortions in active nematics \cite{houston2023active} to show that not only propulsion but also rotation are generic properties of colloidal inclusions. Further, we have demonstrated how the direction and magnitude of these non-equilibrium effects can be controlled through either anchoring conditions or colloidal geometry. The later is illustrated for a form of chiral cog and our predictions for the direction of persistent rotation match those of recent experiments \cite{ray2023}. Our results allow us to go beyond predicting the existence of persistent rotation to also optimise the cog geometry for rotation by maximising the net active torque. 

It is worth emphasising that although the rotational effects we predict seem similar to those reported in bacterial baths \cite{di2010bacterial,sokolov2010swimming}, and even share the property of becoming smoother with increasing teeth number \cite{sokolov2010swimming}, their origin is not the same. The previously observed bacterial ratchets resulted from rectification of the swimming direction of the bacteria to produce coherent motion from an incoherent, isotropic environment and the mechanism for generating rotation is unchanged by rotation of the cog. The orientational order of a nematic gives meaning to the orientation of the cog relative to a far-field alignment and it was by no means guaranteed that the ratchet behaviour would be replicable in this context, particularly given our observation of orientation-dependent handedness \cite{efrati2014orientation} of the nematic distortions that drive the rotation of the cog. It is gratifying then that these effects melt away with increasing tooth number, allowing the active ratchet effect to be recovered.

Our work paves the way to using colloids to induce desired dynamical responses in active nematics, including as work-extracting micromachines \cite{zhang2021autonomous,lv2022nano}. A natural continuation would be to complement the far-field results presented here with a theoretical understanding of how the motion of active nematic defects is rectified by colloidal surfaces \cite{rivas2020driven,ray2023}, as this is key to using inclusions to temper active nematic turbulence \cite{wensink2012meso,alert2022active}.

This work was supported by the UK EPSRC through Grant No.~EP/N509796/1. 

\bibliography{ConformalCogs.bib}

\section{Appendix}
\label{sec:appendix}

\subsection{Free energy for disc colloids}
\label{subsec:disc_energy}

Consider a nematic texture defined by a harmonic director angle
\begin{equation}
    \phi=\frak{I}\left\lbrace\ln\Phi\right\rbrace+A\frac{\ln(\sqrt{z\bar{z}}/R)}{\ln(a/R)}, \qquad \Phi=\prod_j(z-z_j)^{s_j},
\end{equation}
where the $a$ is the colloid radius, $R$ is a large length scale at which uniform alignment is recovered and $A$ is the strength of the monopole term appropriate to enforce boundary conditions. The $z_j$ denote the positions of topological defects of charge $s_j$, both in the bulk and image defects in the interior of the colloid. The Frank free energy density is given by
\begin{equation}
\begin{split}
    &f=4\partial_z\phi\partial_{\bar{z}}\phi=\left(\frac{A}{|z|\ln(a/R)}\right)^2\\
    &+\frac{\text{i}A}{|z|^2\ln(a/R)}\left[\bar{z}\sum_j s_j\frac{z-z_j}{|z-z_j|^2}-z\sum_j s_j\frac{\bar{z}-\bar{z}_j}{|z-z_j|^2}\right]\\
    &+\sum_{i,j}\frac{s_is_j(\bar{z}-\bar{z}_i)(z-z_j)}{|z-z_i|^2|z-z_j|^2}.
\end{split}
\end{equation}
Rewriting this in polar coordinates we have
\begin{equation}
    \begin{split}
        f&=\left(\frac{A}{r\ln(a/R)}\right)^2+\frac{2A}{r^2\ln(a/R)}\sum_j\frac{s_j\mathbf{r}\times\mathbf{r}_j}{|\mathbf{r}-\mathbf{r}_j|^2}\\
        &+\sum_{i,j}\frac{s_is_j(\mathbf{r}-\mathbf{r}_i)\cdot(\mathbf{r}-\mathbf{r}_j)}{|\mathbf{r}-\mathbf{r}_i|^2|\mathbf{r}-\mathbf{r}_j|^2}.
    \end{split}
\end{equation}
The second of these terms integrates to zero by symmetry and in what follows it will be possible to enforce $A=0$, which will hold for the minimal energy configuration. We therefore consider only the final term and write
\begin{equation}
        F=\frac{K}{2}\int\text{d}^2\mathbf{r}\left[\sum_{i,j\neq 0}\frac{s_is_j(\mathbf{r}-\mathbf{r}_i)\cdot(\mathbf{r}-\mathbf{r}_j)}{|\mathbf{r}-\mathbf{r}_i|^2|\mathbf{r}-\mathbf{r}_j|^2}\right].
    \label{eq:FreeEnergyInitial}
\end{equation}

To perform the angular integral we use \cite[eq. 2.559(2)]{gradshteyn2014table}

\vspace{1cm}

\begin{widetext}
    \begin{equation}
    \begin{split}
        &\int\frac{A+B\cos x+C\sin x}{(a_1+b_1\cos x+c_1\sin x)(a_2+b_2\cos x+c_2\sin x)}\text{d}x=A_0\ln\frac{a_1+b_1\cos x+c_1\sin x}{a_2+b_2\cos x+c_2\sin x}+A_1\int\frac{\text{d}x}{a_1+b_1\cos x+c_1\sin x}\\
        &\hspace{9cm}+A_2\int\frac{\text{d}x}{a_2+b_2\cos x+c_2\sin x},
        \end{split}
\end{equation}
where 
\begin{equation}
    \begin{split}
        &A_0 = {\cal N}^{-1} \begin{vmatrix}
        A & B & C\\
        a_1 & b_1 & c_1\\
        a_2 & b_2 & c_2
\end{vmatrix}, \qquad
A_1 = {\cal N}^{-1} \begin{vmatrix}
\begin{vmatrix}
B & C\\
b_1 & c_1
\end{vmatrix} & \begin{vmatrix}
A & C\\
a_1 & c_1
\end{vmatrix} & \begin{vmatrix}
B & A\\
b_1 & a_1
\end{vmatrix}\\
a_1 & b_1 & c_1\\
a_2 & b_2 & c_2
\end{vmatrix}, \qquad 
A_2 = {\cal N}^{-1} \begin{vmatrix}
\begin{vmatrix}
C & B\\
c_1 & b_1
\end{vmatrix} & \begin{vmatrix}
C & A\\
c_1 & a_1
\end{vmatrix} & \begin{vmatrix}
A & B\\
a_1 & b_1
\end{vmatrix}\\
a_1 & b_1 & c_1\\
a_2 & b_2 & c_2
\end{vmatrix}, \\
& \mathcal{N}=\begin{vmatrix}
a_1 & b_1\\
a_2 & b_2
\end{vmatrix}^2-\begin{vmatrix}
b_1 & c_1\\
b_2 & c_2
\end{vmatrix}^2+\begin{vmatrix}
c_1 & a_1\\
c_2 & a_2
\end{vmatrix}^2
    \end{split}
\end{equation}
\end{widetext}
to show that
\begin{equation}
    \begin{split}
       &\int_{-\pi}^{\pi}\frac{(\mathbf{r}-\mathbf{r}_i)\cdot(\mathbf{r}-\mathbf{r}_j)}{|\mathbf{r}-\mathbf{r}_i|^2|\mathbf{r}-\mathbf{r}_j|^2}\text{d}\theta\\
       &=\frac{r^2-r_ir_j\cos(\theta_i-\theta_j)}{r^4+r_i^2r_j^2-2r^2r_ir_j\cos(\theta_i-\theta_j)}\\
       &\qquad\pi\left[\text{sign}(r-r_i)+\text{sign}(r-r_j)\right].
       \label{eq:AngularIntegral}
    \end{split}
\end{equation}

\subsubsection{Dipolar colloids}
We now consider the particular configurations of interest. For a disc of radius $a$ inducing a dipolar distortion with an anchoring angle of $\alpha$ the director angle is given by
\begin{equation}
    \phi=\frak{I}\left\lbrace\ln\left[\frac{z^2}{(z-z_d)(z-\frac{a^2}{\bar{z}_d})}\right]\right\rbrace+(\theta_d+\alpha)\frac{\ln(\sqrt{z\bar{z}}/R)}{\ln(a/R)},
    \label{eq:DipoleDirectorAngleAppendix}
\end{equation}
that is a bulk $-1$ defect at $z_d=r_d \text{e}^{i\theta_d}$, an image $-1$ defect within the colloid and a $+2$ defect at the origin. Clearly the energy is minimised by the choice $\theta_d=-\alpha$ such that the angle of the bulk defect is determined by the anchoring condition. This angular position of the bulk defect is the relevant part for determining the direction of self-propulsion in an active system, but for completeness we determine the radial position.

Using \eqref{eq:DipoleDirectorAngleAppendix} in \eqref{eq:FreeEnergyInitial} and performing the angular integration gives
\begin{equation}
    \begin{split}
        F&=\pi K\int\text{d}rr\left[-\frac{2}{r^2}+\frac{1}{r^2-a^2}+\frac{1}{r^2-a^4/r_d^2}\right.\\
        &\left.+\left(-\frac{2}{r^2}+\frac{1}{r^2-a^2}+\frac{1}{r^2-r_d^2}\right)\text{sign}(r-r_d)\right].
    \end{split}
\end{equation}
To regularise the integral we integrate from $a+\epsilon$ to $R$, excluding where necessary an annulus of width $2\epsilon$ centred on $r_d$. Doing this and expanding the resulting logarithms to first order in $\epsilon$ and $1/R$ yields
\begin{equation}
    \begin{split}
        F&=\pi K\left[\frac{a^2+r_d^2}{a^3-ar_d^2}\epsilon-\frac{2a^2+a^4/r_d^2+r_d^2}{2R^2}-\ln(\epsilon)\right.\\
        &\left.+\ln\left(\frac{r_d^4}{2a(r_d^2-a^2)}\right)\right],
    \end{split}
\end{equation}
with the first two terms arbitrarily small in the limits of small $\epsilon$ and large $R$, the third term the isolated divergent contribution and the final term minimised by $r_d=\sqrt{2}a$.

\subsubsection{Quadrupolar colloids}
Again considering a disc of radius $a$ with anchoring angle $\alpha$ but now inducing a quadrupole distortion, the director angle is given by
\begin{equation}
    \begin{split}
        \phi&=\frak{I}\left\lbrace\ln\left[\frac{z^2}{\sqrt{(z-z_{d1})(z-a^2/\bar{z}_{d1})(z-z_{d2})(z-a^2/\bar{z}_{d2})}}\right]\right\rbrace\\
        &+\left(\frac{\theta_{d1}+\theta_{d2}}{2}+\alpha\right)\frac{\ln(\sqrt{z\bar{z}}/R)}{\ln(a/R)},
    \label{eq:QuadrupoleDirectorAngleAppendix}
    \end{split}
\end{equation}
where the $-1$ bulk defect of the dipolar distortion has been split into two $-1/2$ defects, each with image defects within the colloid. As the free energy is symmetric under exchange of $r_{d1}$ and $r_{d2}$ the global minimiser must occur when $r_{d1}=r_{d2}$. Similarly, from \eqref{eq:AngularIntegral} we can see that the angular positions of the defects will only appear in the combination $\cos(\theta_{d1}-\theta_{d2})$ and so minimising gives the condition $\theta_{d1}-\theta_{d2}=\pi$. It is natural that these restrictions are required to take \eqref{eq:QuadrupoleDirectorAngleAppendix} into the form of the global energy minimiser as they also ensure that the leading order distortion is a quadrupole rather than a dipole. Taking this angular condition in conjunction with the constraint that the monopole term vanishes, that is $\theta_{d1}+\theta_{d2}=-2\alpha$, gives that the two defects sit at $-\alpha\pm\pi/2$.

With these simplifications in place the free energy is given by
\begin{equation}
    \begin{split}
        F&=\pi K\int\text{d}r r\left[\frac{a^4(r^4r_d^4+a^4r^4-2a^8}{r^2(a^4-r^4)(a^8-r^4r_d^4)}\right.\\
        &\left.+\frac{r^4r_d^4+a^4(r^4-2r_d^4)}{r^2(r^4-a^4)(r^4-r_d^4)}\text{sign}(r-r_d)\right].
    \end{split}
\end{equation}
Integrating over $r$ while employing the same regularisation proceedure as before and then expanding to lowest order in $\epsilon$ and $1/R$ we find
\begin{equation}
    \begin{split}
        F&=\frac{\pi K}{2}\left[\frac{2(a^4+r_d^4)}{a(a^4-r_d^4)}\epsilon-\frac{(a^4+r_d^4)^2}{2r_d^4R^4}-\ln\epsilon\right.\\
        &\left.+\ln\left(\frac{r_d^7}{4a^2(r_d^4-a^4)}\right)\right].
    \end{split}
\end{equation}
As before the only relevant term is the final one and it is minimised by $r_d=(7/3)^{1/4}a$.

\subsection{Schwarz-Christoffel Maps}
\label{subsec:Schwarz-Christoffel}

Schwarz-Christoffel transformations are predicated on the notion that the desired mapping has a derivative expressible as the product of some set of canonical functions, that is \cite{driscoll2002schwarz}
\begin{equation}
    f'(z)=\prod f_k(z).
    \label{eq:SCMapForm}
\end{equation}
It naturally follows that $\text{arg}f'=\Sigma\text{arg}f_k$ and so by choosing the $f_k$ appropriately such that each is a step function we can contrive to make $\text{arg}f'$ a piecewise constant function with prescribed jumps and hence map the real axis to a polygon. Through additional modifications we can generate our desired transformations from the exterior of the disc to the exterior of polygons.

\sloppy To make this precise, let us begin by setting out our notion for polygons. We define a polygonal colloid through a piecewise linear curve $\Gamma$ comprised of $n$ vertices $w_1,\dots,w_n$ and bounding an interior region $P$. The vertices have interior angles $\mu_1\pi,\dots,\mu_n\pi$. The corresponding exterior angles are $\nu_k\pi$, with $\nu_k=1-\mu_k$, and for $\Gamma$ to form a closed curve we require $\sum_k\nu_k=2$.

\begin{figure}
    \centering
    \includegraphics[width=0.95\linewidth, trim = 0 0 0 0, clip, angle = 0, origin = c]{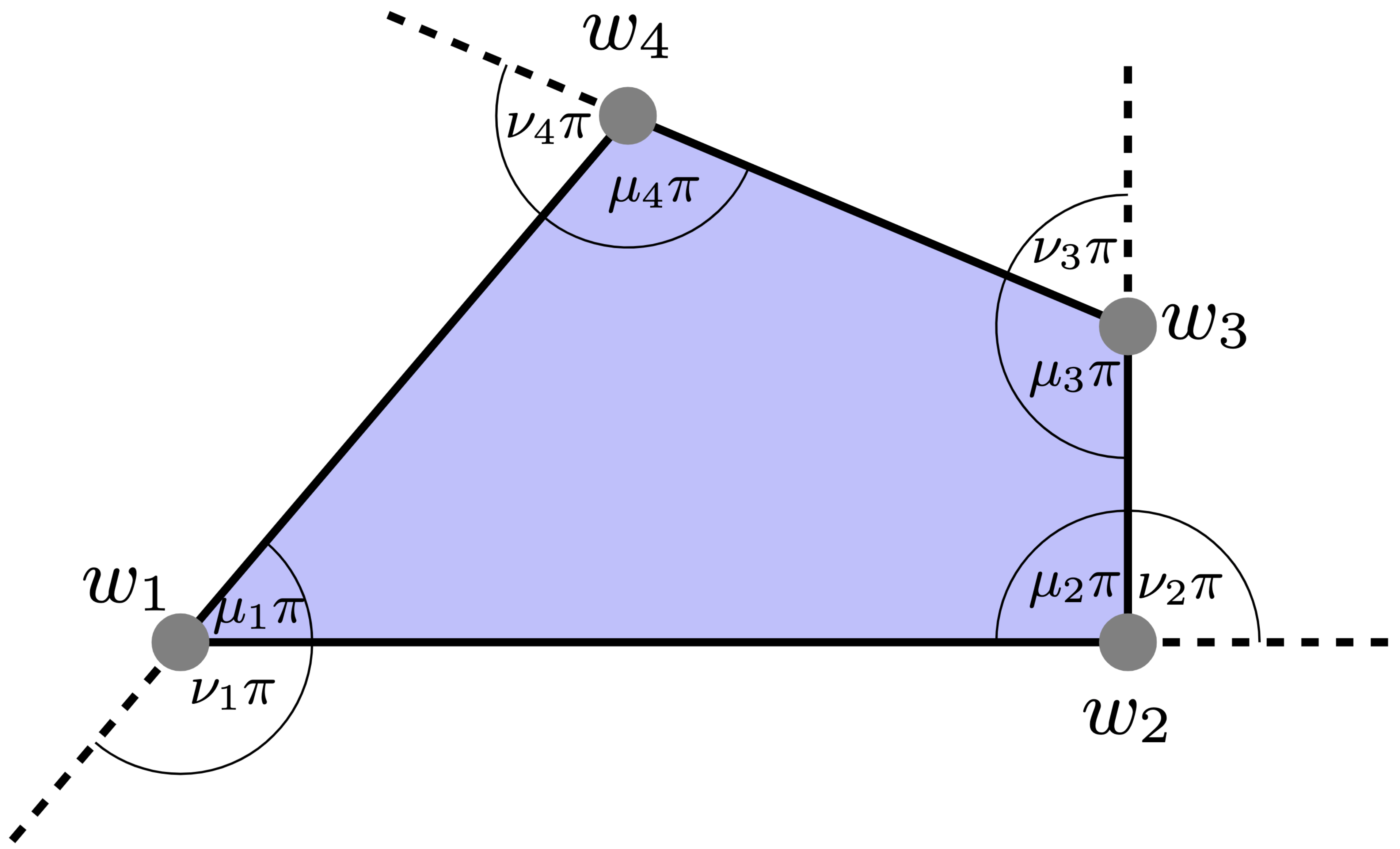}
    \caption[The notation for labelling polygons.]{The notation for labelling polygons. The vertices $w_i$ have interior angles $\mu_i\pi$ and exterior angles $\nu_i\pi$, with $\nu_i=1-\mu_i$.
    }
    \label{fig:my_label}
\end{figure}

To determine the form of $f_k(z)$ in \eqref{eq:SCMapForm} we note that $(z-z_k)^{-\nu_k}$ has a constant argument on $\mathbb{R}$ save for a jump of $\nu_k\pi$ at $z_k$ and is analytic in the upper half-plane $H^+$. It follows that a conformal transformation that maps $\mathbb{R}$ onto $\Gamma$ and $H^+$ onto $P$ is provided by
\begin{equation}
    f(z)=A+C\int^z\prod_{k=1}^n(z'-z_k)^{-\nu_k}\text{d}z'.
    \label{eq:SCH+P}
\end{equation}
The $z_k$ are the prevertices, preimages of the vertices under the map such that $w_k=f(z_k)$. That these prevertices are well-defined is a consequence of the Carath\'{e}odory–Osgood theorem \cite{driscoll2002schwarz,henrici1974appl}. The complex constants $A$ and $C$ allow for scaling, translation and rotation of the image. Together with the prevertices they form a set of parameters which are overdetermined to the tune of the three degrees of freedom inherent in Riemann's mapping theorem \cite{ahlfors1979introduction}. The lower limit of integration is left unspecified as it only provides a constant that may be absorbed into $A$. The proof that \eqref{eq:SCH+P} truly is an analytic function as required is based upon showing that
\begin{equation}
    \mathcal{F}(z)=\frac{f''(z)}{f'(z)}-\sum_k\frac{\nu_k}{z-z_k}
\end{equation}
is an entire function and may be found in \cite{driscoll2002schwarz,nehari2012conformal,MyThesis}.

There are two modifications we must make to \eqref{eq:SCH+P} that we now consider in turn. The first is to change the domain of the map from the upper half-plane to the unit disc, $D$. This is achieved by composition of maps and application of the chain rule. For $h(z):H^+\to P$ and $g(z):\Sigma\to H^+$ we have $f(z)=h(g(z)):\Sigma\to P$ with \cite{driscoll2002schwarz}
\begin{equation}
    \begin{split}
        f'(z)&=h'(g(z))g'(z)\\
        &=Cg'(z)\prod_k\left[g(z)-g(z_k)\right]^{-\nu_k}.
    \end{split}
\end{equation}
A map from the disc to the upper half-plane is provided by the M\"{o}bius transformation $g(z)=\text{i}\frac{1+z}{1-z}$, giving
\begin{equation}
    f'(z)=C\frac{2\text{i}}{(1-z)^2}\prod_k\left[\frac{2\text{i}(z-z_k)}{(1-z)(1-z_k)}\right]^{-\nu_k}.
\end{equation}
Using that $\sum_k\nu_k=2$ and absorbing various constants into $C$ we integrate to obtain
\begin{equation}
    f:D\to P,\qquad f(z)=A+C\int^z\prod_k\left(1-\frac{z'}{z_k}\right)^{-\nu_k}\text{d}z'.
\end{equation}

Next we must consider mapping to the exterior region $P'$. The first change that this necessitates is reversing the sign of the exterior angles. In the same manner as before this leads us to
\begin{equation}
    \mathcal{F}(z)=\frac{f''(z)}{f'(z)}-\sum_k\frac{\nu_k}{z-z_k}
    \label{eq:ExteriorMapAnalyticF}
\end{equation}
as an entire function. The difference now is that since the image of $f$ contains the point at infinity it no longer follows that $\mathcal{F}(z)$ is constant. Taking $f(0)=\infty$ implies that $f$ has a simple pole there and so at the origin $f''/f'=-2/z$ plus some analytic part. We therefore modify \eqref{eq:ExteriorMapAnalyticF} to give
\begin{equation}
    \mathcal{F}_2(z)=\frac{f''(z)}{f'(z)}-\sum_k\frac{\nu_k}{z-z_k}+\frac{2}{z}=0.
\end{equation}
Integrating we find
\begin{equation}
    f:D\to P', \qquad f(z)=A+C\int^z z'^{-2}\prod_k\left(1-\frac{z'}{z_k}\right)^{\nu_k}\text{d}z'.
    \label{eq:SCDP'}
\end{equation}
Finally, we may make the replacement $z\to1/z$ in \eqref{eq:SCDP'} to attain our desired conformal map from the exterior of the disc to the exterior of the polygon. 

Having attained this general expression we now derive the particular mappings appropriate to the polygonal cogs that we consider in this paper. Our cogs are formed by attaching right-angled triangles with base angle $\gamma\pi$ to each side of a regular $n$-sided polygon and we can use this rotational symmetry to simplify the integral. We take the prevertices of the base $n$-gon to be the $n^{\text{th}}$ roots of unity and indeed to be fixed points of the map. Due to the cyclic nature of the roots of unity $\prod_k\left(1-\frac{z'}{z_k}\right)=1-z'^n$. A similar relation holds for the prevertices of the tips of the cog teeth, with these being rotated from the roots of unity by an angle $\chi$ which depends on both $n$ and $\gamma$ and must be determined numerically. In addition we can set $A=0$ and determine $C$ by requiring $f(1)=1$ such that
\begin{equation}
     f(z)=\frac{\int^{z}z'^{-2}(1-z'^n)^{\frac{4-n}{2n}-\gamma}\left(1-\frac{z'^n}{\text{e}^{\text{i}n\chi}}\right)^{\frac{1}{2}+\gamma}\text{d}z'}{\int^{1}z'^{-2}(1-z'^n)^{\frac{4-n}{2n}-\gamma}\left(1-\frac{z'^n}{\text{e}^{\text{i}n\chi}}\right)^{\frac{1}{2}+\gamma}\text{d}z'}.
     \label{eq:SCCogInitial}
\end{equation}
Making the substitution $t=z'^n/z^n$ brings the numerator to the form
\begin{equation}
    \frac{1}{nz}\int^1t^{-\frac{1}{n}-1}\left(1-z^nt\right)^{\frac{4-n}{2n}-\gamma}\left(1-\frac{z^n}{\text{e}^{\text{i}n\chi}}t\right)^{\frac{1}{2}+\gamma}\text{d}t.
\end{equation}
Comparing with Picard's single integral representation of Appell's $F_1$ function \cite[Eq.~16.15.1]{NIST:DLMF}
\begin{equation}
\begin{split}
    &F_1(a,b_1,b_2,c;x,y)=\\
    &\frac{\Gamma(c)}{\Gamma(a)\Gamma(c-a)}\int_0^1t^{a-1}(1-t)^{c-a-1}(1-xt)^{-b_1}(1-yt)^{-b_2}\text{d}t,
\end{split}
\label{eq:Picard}
\end{equation}
we find that the numerator of \eqref{eq:SCCogInitial} is given by
\begin{equation}
    \frac{1}{nz}\frac{\Gamma(-\frac{1}{n})}{\Gamma(1-\frac{1}{n})}F_1\left(-\frac{1}{n},\frac{n-4}{2n}+\gamma,-\frac{1}{2}-\gamma;z^n,\frac{z^n}{\text{e}^{\text{i}n\chi}}\right).
    \label{eq:SCCogNumerator}
\end{equation}
We note that the representation in \eqref{eq:Picard} is only valid for $\frak{R}a>0$, whereas here $a=-1/n$, leading to a singularity at $t=0$. This singularity can be handled via adaptation of the integration contour, introducing a prefactor into \eqref{eq:SCCogNumerator}. We leave this prefactor undetermined as it is common to both the numerator and denominator of \eqref{eq:SCCogInitial} and hence does not feature in the desired mapping $f(z)$.

For the denominator we use \cite[Eq.~5.10.10]{bateman1953higher}
\begin{equation}
    \begin{split}
        F_1(a,b_1,b_2,c;1,z)&={}_2F_1(a,b_1;c;1){}_2F_1(a,b_2;c-b_1;z)\\
        &=\frac{\Gamma\left(c\right)\Gamma\left(c-a-b_1\right)}{\Gamma\left(c-a\right)\Gamma\left(c-b_1\right)}{}_2F_1(a,b_2;c-b_1;z),
    \end{split}
\end{equation}
with the same caveat concering a prefactor as for \eqref{eq:SCCogNumerator}. Replacing $z$ with $1/z$ and including rotations of the image polygon by an angle $\psi$ we  reach the conformal transformation from the exterior of the disc to the exterior of a cog

\vspace{1cm}

\begin{widetext}
\begin{equation}
    f(z)=\frac{\Gamma\left(\frac{1}{2}+\frac{1}{n}-\gamma\right)zF_1\left(-\frac{1}{n},\gamma+\frac{n-4}{2n},-\frac{1}{2}-\gamma,1-\frac{1}{n};\text{e}^{\text{i}n\psi}z^{-n},\text{e}^{\text{i}n(\psi+\chi)}z^{-n}\right)}{\Gamma\left(1-\frac{1}{n}\right)\Gamma\left(\frac{1}{2}+\frac{2}{n}-\gamma\right){}_2F_1\left(-\frac{1}{n},-\frac{1}{2}-\gamma,\frac{1}{2}+\frac{1}{n}-\gamma,\text{e}^{\text{i}n\chi}\right)}.
\end{equation}
\end{widetext}

\end{document}